# Emotions in Pervasive Computing Environments

**Nevin VUNKA JUNGUM[1] and Éric LAURENT[2]**

**[1] Computer Science and Engineering Department,
University of Mauritius
Réduit, Mauritius**

**[2] Laboratoire de Psychologie, ENACT-MCA,
University of Franche-Comté
Besançon, France**

## Abstract

The ability of an intelligent environment to connect and adapt to real internal sates, needs and behaviors' meaning of humans can be made possible by considering users' emotional states as contextual parameters. In this paper, we build on enactive psychology and investigate the incorporation of emotions in pervasive systems. We define emotions, and discuss the coding of emotional human markers by smart environments. In addition, we compare some existing works and identify how emotions can be detected and modeled by a pervasive system in order to enhance its service and response to users. Finally, we analyze closely one XML-based language for representing and annotating emotions known as EARL and raise two important issues which pertain to emotion representation and modeling in XML-based languages.

## 1. Introduction

Problems or scenarios in real-life and computer processing logic can be represented using mathematical equations, whereby objects are represented using variables and constants and relationships using operators. Consider the following equation that represents a specific situation/scenario:

$$y = x^2 + x + 1 \qquad \text{(Eq. 1)}$$

When $x = 1$,

$$y = 1^2 + 1 + 1 = 3 \text{ (using mathematical logic; this is what computers use)}.$$

Now, in real-life:

$$y = x^2 + x + 1 \qquad \text{(Eq. 1)}$$

When $x = 1$,

$y$ is either **3** or **something else**, that is, **not 3** (using human reasoning; this is what human used).

Because of emotions, people tend to add an additional variable which represent the state of emotion. And this changes the equation to:

$$y = x^2 + x + 1 + \boldsymbol{\mu}, \qquad \text{(Eq. 2)}$$

where $\boldsymbol{\mu}$ is a variable whose value varies with the state of emotion. Thus possibly leading to a non-logical response from humans; this is how even the most intelligent person on earth can make "errors", or say, produce a variable behavior relatively to what could be predicted on the basis of a strictly rational norm; whereas the less powerful and oldest computer will never since its calculations are based entirely on logic without any emotion. This view on emotions corresponds to the classical sketch where information processing abilities are decreased. Therefore emotions are seen as negative processes that tend to diminish the power of the system. Though opinions on emotions in psychology have changed – because emotion is also considered as a positive and adaptive process – we will try to show that current models of pervasive computing systems deal poorly with emotions, including when the latter is restricted to its negative influences on cognition.

The fundamental aim of a non-simulated real-life pervasive computing environment is to support users in the environment, by providing personalized services and eliminate the user's thought that he/she is dealing with computers to accomplish his/her task. How can a system, whose environment is based on mathematical logic, be used to support another system which is based on human logic/reasoning which is influenced by emotions?

Is it an important issue? Yes it is. Consider the following scenario. In a smart building of tomorrow, the environment is supported with the deployment of a pervasive computing system. Everything is automated and access to resources is autonomous and invisible, say, users are identified using their mobile phones. Previously, before becoming a smart building, the old Mr. Harry, the storekeeper, used to control access to equipments used for maintenance like hammers and all that. However, due to advance in technology, the service of Mr. Harry was no more required. Access to resources like equipments was controlled using the pervasive computing system. Jack and Paul were two colleagues working for the maintenance department. The relation between the two was not that good due to conflict of interest at workplace. One day, like any other day, they were busy abusing and arguing with each other. But this time things were getting worst. In a fit of rage, without thinking about the consequence, Jack went to the store, since he was identified as a maintenance staff by the system, so he was allowed access and he took the hammer and went back to Paul, and he hit the latter on his head with the hammer. Imagine if the pervasive computing system was not introduced, Mr. Harry, the storekeeper, would have still been there, and when he would have seen Jack very angry entering the store taking a hammer to go and fight with Paul, surely, without hesitation he would have stopped him. Jack's access to the resources would have been denied.

So why has the Pervasive Computing System failed? The reason by this time is crystal clear. The system did not take any emotional state into consideration as Mr. Harry did. If ever the middleware on which the pervasive computing system deployed were built had considered emotion capture, emotion processing, emotion identification, hence, emotion management, then the system would have definitely denied Jack from accessing that resource. The issue of emotion handling is of fundamental importance for real-life non-simulated pervasive computing system and has been acknowledged in various works [2][3][4][5][37][38]. Systems that considered emotions are referred as emotion-oriented computing or Affective Computing systems; words coined by Rosalind Picard [1] in 1997.

If a Pervasive Computing System does take the emotion of people into consideration, then the gap between mathematical reasoning of computers and human reasoning will be shorter, and hence the Pervasive Computing System will be in a much better position to support users in the environment in an effective way.

The paper is structured as follows: Section 2 covers issues pertaining to cognition, emotion and behavioral variability in humans. These issues are discussed in the framework of two trends (i.e., classical cognitive psychology, enactive psychology) and a summary of current knowledge about emotional markers is provided. Section 3 analyzes existing pervasive computing systems that take emotions into consideration based on fundamental emotional sources. Detection or emotions capture and representation, and modeling are discussed in Section 4, and finally Section 5 concludes with some words for future works.

## 2. Cognition, emotion, and behavioral variability in humans

### 2.1 Cognition and the problem of human complexity

Cognitive psychology, dealing with human information processing has developed since the end of World War two, heavily influenced by information theory – emerging at that time under the influence of the work of Shannon and Weaver in telecommunication [61][62]. The linear aspect of the relations between the speed of a system's processing and the amount of information to process is a key feature allowing to predict the system's behavior in the framework of a simple and bright fashion. Hick [51] applied to man the theory in order to predict Reaction Time (RT) as a function of the number of possible choices (N) in a choice RT task:

$$RT = b + a.\log2(N) \qquad\qquad \text{(Eq. 3)}$$

Hick's law was one of the first formal account of human cognition, based on the idea that human being could be assimilated to an information processing system highly predictable on the basis of the knowledge of the amount of information it has to process [$\log_2$(N)] expressed in bits of information. It is noteworthy that in this case, the whole behavioral variability (indexed by the reaction time profile) is predicted by a simple parameter: the number of bits of information easily derived by researchers from the number of empirical possible choices. Moreover, and most importantly for the development of cognitive psychology during the 20th century, these early developments (see also the works of Fitts [52]) influenced contemporary classical cognitive theories that explicitly or implicitly conceive a rational agent able to make computations on abstract symbols. The computational metaphor has been associated to a dissociation between research on cognition on the one hand and research on emotions and affects on the other hand. As a result, cognition was for several decades considered independently of affective, cultural, social and physiological reality [70]. The problem of psychological variability was then reduced more or less implicitly to a problem of "informational" complexity, following Hick's work and other influential seminal research in the field.

## 2.2 What is human emotion? How important is it? A view from "enactive" psychology

The view according to which humans are able to process information in a rather stable way is questioned by research carried out in the psychology of emotions, even if the two fields (i.e., cognition, emotion) were significantly interacting only from recent times. This is an important issue to us since it concerns the predictability of the agent's needs, perception, decisions, and behaviors by smart environments. For instance, the more stable the agent's needs, the easier the programming of Pervasive Computing Environments. In our framework, emotion understanding is a key to predict human cognition, and more generally psychological variability.

*Emotion*, as mood, is a specific type of affective state. An affective state is generally considered as a form of "representation" (e.g., behavioral and expressive, cognitive, neurological, physiological, experiential) of personal value [59]. This 'valence' emerges either as a function of a specific and identifiable object or situation, or as a function of a more diffuse scenario. In the first case we talk about "emotions", whereas in the second one the term "mood" is coined. While dealing with emotions, researchers generally underline the importance of appraisal processes in the understanding of emotions' emergence and their consequences. More specifically, *emotion can be defined as* a complex syndrome expressing through a variety of forms such as activation of the autonomic nervous system, disposition to engage in certain actions or social roles, facial expression, and subjective experience [64], but also such as gesture, movements, and vocal production [45]. Therefore, emotion seems to penetrate and influence a series of biological, behavioral, and phenomenal processes. We argue that this is a chance for specialists interested in Pervasive Computing Environments. Indeed, this variety of expressions could allow us to infer internal processes and the current needs of the organism from the pick-up of "emotional markers". Those markers are thought of as both input data for smart environments and the source of online environment adaptation. This line of thought is strongly influenced by one current approach called the *"enactive approach to cognition"*, which conceives that cognition is *"in action"*, that is, almost never static, always moderated or driven by current embodied goals of the organism [57][58]. These goals may have been embodied on different time scales (i.e., phylogeny, ontogeny, scale of neuronal micro-dynamics) [90]. In this framework, the analysis of emotion has been seen as a means to understand and predict how cognition may be changed on the basis of an "online view" on the cognition's dynamics. This "enactive" point of view gather both recent perspectives on dynamic cognition and more restricted interests developed for interactions between cognition and emotion in the field of emotion research.

## 2.3 Tracing back emotional states and psychological needs from the detection of emotional markers

"*Recent years witnessed a revival of research interest in the interplay between cognition and emotion –a subject that stimulated much debate and discussion among psychologists in the nineteenth century, but was shunned throughout the twentieth*" (p. 3). As put by Eich and Schooler [63], the study of emotion was for a long time split from the study of cognition. When you design intelligent environments, you need them to be well-fitted to cognitive states, desires and needs of people. This is important because different underlying psychological processes generally produce *different uses of the environment*. What we know today about human cognition is that this is dynamic, that is, changing over time, and strongly influenced by other psychological or biological processes [57][58][66][67]. By capturing emotion, a smart system would be able to gather information about the usefulness of different environmental solutions.

Our endeavor in building such smart environments will be based on 1) categorizing different types of emotions corresponding to different states and psychological *"orientations"* (each orientation conducting to search for specific things or services in the environment) 2) to describe the way expression of emotions can be used by others than the affected subject (i.e., other humans, artifacts) to infer affective states of this subject and predict his behavior.

It should be noted that there is no such thing as a universal consensus on an exhaustive taxonomy of basic human emotions. However, as noted by Niedenthal and Halberstadt [71], many authors hypothesize that "affective space is organized according to some number of specific or basic emotions" (p. 174). In Table 1, we present a non-exhaustive classification of 6 basic emotions and their behavioral signification, based on MacLean [72]. The leading idea is that emotions are responses developed as a consequence of vital problems faced during the evolution. These emotional responses are thought of as means to motivate adaptive behaviors that respond to basic problems.

Table 1 Example of basic emotions classification based on MacLean (1993)

| Emotion | Motivated behavior |
|---------|-------------------|
| Desire | Searching |
| Anger | Aggressive |
| Fear | Protective |
| Sadness | Dejected |
| Joy | Gratulant (triumphant) |
| Affection | Caressive |

In this perspective, it sounds clear that emotional states are strongly linked to internal needs and direct behaviors towards specific forms of usage of the environment.

Emotions influence a series of behaviors in humans. Emotions are embodied in facial expressions [24][25][39][40][41][45][48][74][91], voice or speech and sound [30][31][32][33][34][35][36][45][46][48], hand gestures [26][27][42][43], body movements [28][29][44][45][47] and specifically kinetic and kinematic features of movements [73], and linguistic lexicon [54][74][75]. It has been established that humans are able to extract information about emotional states through these various source of embodiment [45]. It seems that this process is partially adapted to the interpretation of emotional states of individuals belonging to other species, such as horses for instance [69]. The ability to extract emotional information generally increases with development [49], but the advancement in age can be limiting at some point, especially for inferring emotions from lexical stimuli [74]. Actually, it seems that facial information processing is less affected by ageing [74]. The robustness of facial emotions processing throughout life ages may be a first clue in the process of furnishing different markers' weights to smart environments. This option has got further justifications, which are grounded in developmental and neuropsychological studies that show that 1) processing of facial expressions, especially from features located in the region of the eyes, has got a high value from an evolutionary perspective and is arising very early (i.e., present in some forms at 3 months) during human baby development [92] and 2) in contrast, interpretation of emotions seems to be diminished or impossible respectively for ADHD or schizophrenic patients and this is strongly linked to the inability of such patients to correctly structure their search for information in faces, as studies on eye movement recording and visual search in patients suggest [93]. However, building the coding of emotional information on facial expression capturing, implies that smart environments have a sufficient number of quality interfaces for tracking facial features whatever the moment of the individuals' evolution in the environment. Horstmann and Ansorge have shown very recently [98] that dynamic cues are important for facial expression recognition. In addition to static features, facial *movements* are rich cues for identifying these expressions. Moreover, in some situations, they can even become necessary for a correct identification. For example, in their first experiment, the authors found that humans were efficient for searching a dynamic angry face among dynamic friendly faces, but inefficient in searching when static faces were shown. In subsequent experiments, authors showed that the degree of movement is critical for discriminating the nature of emotions: dynamic angry faces are associated to a stronger movement signal than dynamic friendly faces.

Beyond facial cues, language conveys a series of stimulations that express emotions, and by this means information about human's relation to his environment. In social contexts, where people is talking or writing to other persons, emotional events can be extracted from the analysis of linguistic productions. To this end, we can analyze linguistic labels used in psychological studies in which participants use labels to describe felt (personally) or recognized (in others) emotions. In Table 2 we present examples of linguistic labels associated to different emotions distinguished in a study in which participants had to describe types of emotions from various context (everyday, emotion felt while listening to music, emotion perceived while listening to music) [68].

Table 2 Examples of linguistic markers for various emotions from Zentner, Grandjean and Scherer [68]

| Emotion | Associated lexicon |
|---|---|
| Activation | Disinhibited, excited, active, agitated, energetic, fiery |
| Amazement | Amazed, admiring, fascinated, impressed, goose bumps, thrills |
| Dysphoria | Anxious, anguished, frightened, angry, irritated, nervous, revolted, tense |
| Joy | Joyful, happy, radiant, elated, content |
| Power | Heroic, triumphant, proud, strong |
| Sadness | Sorrowful, depressed, sad |
| Sensuality (desire) | Sensual, desirious, aroused |

One can use linguistic markers of emotions gathered in various life contexts in order to create a knowledge base, which will be provided to the artificial smart system when the human tracked is in communication settings. This base would then serve as reference for interpreting the emotional meaning of discourse.

Voice has also been explored for a long time. In 1872, in his monograph on the expression of emotion, Darwin [99] had already put forward the role of voice for its role in the communication of emotions. As stressed by Banse and Scherer [46] vocal affect communication has got functional value for the "organismic states" through arousal and valence for instance as well as for "interorganismic relationships" through dominance or nurturance for example. Though some emotions are more easily recognizable (i.e., anger, joy, sadness) than others that are generally poorly recognized (e.g., disgust) through vocal expression [100], researchers have suggested that there is little doubt about the existence of vocal variables that signal emotions [46]. These variables are reported in [46] as: "(a) the level, range, and contour of the fundamental frequency (referred to as *F0* […]; it reflects the frequency of the vibration of the vocal folds and is perceived as pitch); (b) the vocal energy (or amplitude, perceived as intensity of the voice); (c) the distribution of energy in the frequency spectrum (particularly the relative energy in the high- vs. low-frequency region, affecting the perception of voice quality or timbre); (d) the location of formants (F1, F2… F$n$, related to the perception of articulation; and (e) a variety of temporal phenomena, including tempo and pausing" (pp. 615-616). The interesting point here is that researchers have identified invariant properties in the production of vocal expressions as a function of the emotional state of human beings. We present these properties in Table 3.

Table 3 Voice physical properties examples as a function of emotion type from Banse & Scherer [46]

| Emotion | Voice physical properties |
|---------|---------------------------|
| Anger | Increases in mean F0, mean energy; increases in F0 variability and in the range of F0 (hot anger); increases in high-frequency energy and downward-directed F0 contours; increase in the rate of articulation |
| Fear | Increases in mean F0, in F0 range, high-frequency energy, and in the rate of articulation |
| Joy | Increases in mean F0, F0 range, F0 variability, and mean energy |
| Sadness | Decreases in mean F0, F0 range, mean energy and downward-directed F0 contours |
| Disgust | No reliable charateristic |

It is noteworthy that some isolated characteristics are found in different types of emotions (e.g., increase in mean F0) and some others are more specific (e.g., decrease in mean F0). Therefore, in order to distinguish between emotions, one should take into account the multiple variables in the same time. As for other markers, we argue that we need to consider a wide range of variables pertaining to various modalities in order to reliably categorize the emotional state of the human subject.

A last type of emotional markers is found in a far less investigated field in the framework of emotion research: body movements. Body movements can constitute an additional source of emotion recognition. This has been studied in the context of dance for instance [47]. It seems that anger, fear, grief, and joy, among others, are especially well recognized by spectators. Time, space, flow and weight are dimensions on which emotional markers can be described for identifying emotional markers related to body movements. The time dimension includes both overall duration and changes in tempo (i.e., the underlying structure of rhythm or flow in the movement). The space dimension refers to a personal space, so that movements can be described spatially in reference to body centre (either in contraction or in expansion relatively to the centre). The flow dimension concerns the shapes of speed and energy curves, and frequency of motion and pause phases. Finally, the weight dimension indexes the amount of tension, and dynamics in movements (measured by the vertical acceleration component). All these components and their associated values on the dimensions are combined and give rise, as a function of emerging values, to a specific movement, which can be associated to an emotion type. Table 4 reports basic body movement properties described on the four above presented dimensions as a function of emotion's types that are generally recognized by spectators.

Table 4 Examples of movement physical properties as a function of emotion type from Camurri, Lagerlöf & Volpe [46]

| Emotion | Body movement physical properties |
|---------|-----------------------------------|
| Anger | Short duration of reaction time<br>Frequent tempo changes, short stops between changes<br>Movements reaching out from body centre<br>Dynamic and high tension in the movement; tension builds up and then 'explodes' |
| Fear | Frequent tempo changes, long stops between changes<br>Movements kept close to body centre<br>Sustained high tension in movements |
| Grief | Long duration of time<br>Few tempo changes, 'smooth tempo'<br>Continuously low tension in the movements |
| Joy | Frequent tempo changes, longer stops between changes<br>Movements reaching out from body centre<br>Dynamic tension in movements; changes between high and low tension |

Other recent experimental data from research investigating directly reaction time and force properties of extension movements in participants submitted to different conditions of emotional induction [73] suggest that emotions modulate premotor reaction time. The exposure to attack tended to shorten premotor reaction time in comparison with all other tested valences. Peak force of movements had a higher value when authors had their subjects exposed to images of erotic couples or to images of mutilation. Further research is needed 1) to more thoroughly investigate the specificity of these motor responses, and 2) in order to know to what extent humans or artificial agents are able to recognize these properties in everyday life. Though reaction time is rather easy to measure when tests are designed, kinetic variables are sometimes hardly reachable, and require heavy (i.e. force platforms) or uncomfortable setups for everyday human living (i.e., electromyographic electrodes for the measure of the intensity of electrical activity of muscles). However this perspective, though new and far less embraced in comparison to previous ones, opens new sources for the coding of emotions and should develop with technological advancement allowing the use of more convenient tools of information capture.

Though we have envisaged a multimodal approach to emotion recognition, it appears that 1) there is little empirical evidence that conjoint multimodal patterns, in the face, the voice, the language, occur 2) "most theories of emotion remain silent as to the mechanism that is assumed to underlie integrated multimodal emotion expression" [45] and recognition. Indeed, even if multiple channels are envisaged, how various inputs are combined into a coherent and general pattern remains poorly explicated despite of the general claim of some basic theories that see emotions as being driven by fixed neuromotor programs. Alternative componential appraisal theories supported by the findings of Scherer and Ellgring [45] assume that coordination between different modalities of emotional expressions is rather variable and is driven by the results of sequential appraisal checks.

In order to design our Pervasive Computing Environment, we will therefore build on several input sources and define their coordination as a function of pragmatic constraints related to human-machine communication's issues. Respective weights of various sources of emotional expression can change as a function of the availability of the different sources of information, whatever the sources. This availability is dependent upon 1) the situation in which humans are immerged (e.g., somebody can be alone or interacting with other humans) 2) the current relationship between the human subject and the pervasive computing environment (e.g., captors may be unable to track their target given the features of the current situation). Moreover some sources such as kinetic characteristics of an individual's actions are less convenient than others because they involve a heavy setup procedure (see Table 5 for an overview of basic capturing conditions and convenience of use of considered techniques as a function of the expressional source).

Table 5 Sources of emotional expression and basic conditions of capture

| Source | Basic conditions of capture | Convenience |
|---|---|---|
| **Face** | Sufficient number of optical devices; sufficient lighting intensity for basic cameras; region of the human's eyes tracked | Good |
| **Language and voice** | Human-human or human-machine oral or written communication | Good |
| **Hand gesture and body movement** (kinematic studies) | Sufficient number of high-frequency optical devices for kinematic coding | Middle |
| **Hand gesture and body movement** (kinetic and electromyographic or "EMG" studies) | Sufficient number of force devices for kinetic and EMG coding | Good for small load cells; bad for force platforms (heavy setup); middle or bad for EMG captors on the human subject (on an everyday basis) |

Until now, we have based our approach on the capture of behavioral or more elementary neuromuscular or physical characteristics of human productions in order to categorize emotional states. In our view, emotional categories or labels alone are insufficient in the perspective of finely adapting smart environments behaviors to human needs. As remarked by researchers who develop dimensional theories of emotions, emotional states are complex emerging states that can be characterized by combining scores on different phenomenal dimensions, such as for example pleasure (or valence), arousal, and dominance [101]; or evaluation-pleasantness, potency-control, activation-arousal, and unpredictability [102]. Though we think that the categorical approach of basic emotions is relevant in order to adapt behaviors of artificial agents as a function of the dominant meaning that can be attributed to an affective state, we will take into account basic principles of dimensional theories. These principles consist in considering the emergence of complex affective states as made of different components. In an analogical way, we will take into account, at another scale, the fact that multiple emotions can emerge in the same time. We will adopt the principle of the weighing of emotions as a function of their respective intensity, that is, as a function of the relative values measured on the diagnostic variables that have been described earlier on the paper.

In conclusion of this section, we want to stress that in contrast with the classic and rationalist view dating back to the ancient Greeks [94], according to which emotion just diminishes the power of rationality, we defend the view that emotions in humans have got functional value in directing cognition and behavior as a function of current individual needs. Based on the *conditions of capture* listed in Table 5 and on *the links between emotions and motivations/needs* presented in Table 1, we will develop, in the following sections, potential solutions for designing smart environments that take into account multidimensional human emotional states. This will be done as a first step in the endeavor of building smart tools capable of online adaptation to human needs through the processing of emotional states.

## 3. Emotions-Oriented Systems

During the last decade researchers have been proposing and some even implemented middlewares [8][9][10][11][12] to enable the Pervasive Computing paradigm. However, the strategic approach for designing the middleware varies. Some of the approaches are as follows: reconfigurable middleware [13], based on techniques for dynamic adaptation of distributed systems [14][18][19], aspect-oriented [12][15], message-oriented [11], dynamic service composition approaches as surveyed in [16], semantic middleware [17], reflective middleware [21] and collaborative content adaptation [20] among others.

However, none of these middleware is suitable if we consider the story of Mr. Harry discussed in the introduction since no multimodal approach is adopted, as we further explain below, when capturing emotions. To address the issue of emotion, various attempts have been made to identify, capture, model and use emotions in pervasive systems. Table 7 below show the analysis of some proposed pervasive systems that took emotions into consideration. No specific selection criteria have been considered during system selection. The models are analyzed based on the five fundamental emotion sources as described above in Section 2.3, namely, facial expressions, body movements, speech and sound, hand gestures and linguistic. The scaling of 1 to 5 is used, where 1 denotes 'No Consideration' and 5 denote 'Strong Consideration'. Refer to Table 6 for more scaling details.

Table 6 Scaling Level Used For System Analysis

| Rating | Degree of consideration |
|--------|-------------------------|
| 1 | Nil / No consideration |
| 2 | Weakly / Less than average |
| 3 | Partially / Medium / Reasonable |
| 4 | Mostly / But not complete |
| 5 | Fully / Strongly / Absolute consideration |

Table 7 Systems' Middleware Analysis

| Emotions Source | System 1 [37] | System 2 [38] | System 3 [6] | System 4 [5] |
|-----------------|---------------|---------------|--------------|--------------|
| Facial Expressions | 1 | 3 | 1 | 5 |
| Body Movements | 1 | 3 | 2 | 1 |
| Speech & Sound | 5 | 3 | 1 | 1 |
| Hand Gestures | 1 | 3 | 1 | 1 |
| Linguistic | 1 | 5 | 1 | 1 |

The authors in [37] investigate the usage of sound captured in a smart environment for elder people to be incorporated in their system as contextual information. They attempt to detect hazards so as to better assist the users of the environment. Each sensor used to determine the sound in the environment use a *Multi-Modal Anxiety Determination* [37] algorithm that determines the anxiety level of the user based on the foreground and background sounds captured. Though the results reported from an experimental phase sounds interesting, however the system focuses only on sound to derive emotion types. For example, if the user does not speak, it would be very difficult, if not impossible, to detect his/her emotional state. Though sound/speech is important and does help in identifying emotions, relying on sound only does not permit flexibility in the system to be confident enough about the actual emotional state of a user at a particular instant.

In [38] the researchers aimed to create a framework for emotion-aware ambient intelligence to facilitate building applications that participate in the emotion interaction. The framework integrates ambient intelligence, affective computing, service-oriented computing, emotion-aware services, emotion ontology, and service ontology. Ontologies are considered to represent emotion knowledge about emotion types, emotion actions, emotion responses, application domains, and the relationships between them. The ontologies can also be used in semantic-based emotional intelligence testing, training, and academic research. They considered the capture of emotions through face, speech and body movements. However, in-depth discussions about how emotions are captured, modeled using ontologies and used to influence system's response were not presented, which thus leave these areas obscured with respect to their proposed framework. Discussions about the structure of emotion in English conversation were also made. They argued that the processing of emotion in English conversation has not been systematically explored and thus explored the structure of emotion in English conversation by studying four linguistic features in English conversation such as lexical choice, syntactic form, prosody and sequential positioning.

At Carnegie Mellon University, the ComSlipper project [6] augments traditional slippers, enabling two people in an intimate relationship to communicate and maintain their emotional connection over long distance. To express emotions such as anxiety, happiness, and sadness, users perform different tactile manipulations on the slippers (e.g. press, tap at a specific rhythm, touch the sensor at the side of slippers, etc.) which ComSlipper recognizes. The remote slipper pair then displays these emotions through changing LED signals, warmth, or vibration. However, users need to learn the different tactile manipulations of slippers and mapping to different emotions, which may not be natural or intuitive for most people. In addition, users need to learn how to interpret messages through LED lights, warmth, or vibration.

The author in [5] proposed an adaptable emotionally rich pervasive computing system to cater for the different ways of expressing emotions by different people in different circumstances. It is argued that the proposed architecture can automatically update its performance to a particular individual by taking into account the user's characteristics and properties. It also takes into consideration the emotions of other surrounding people (family, friends, and work mates) in the environment. However, though results are presented that illustrate the efficiency of the proposed scheme in recognizing the emotion of different people or even the same under different circumstances, the model of emotion determination mechanism is solely based on facial expressions analysis. Other important sources of capturing emotions as described above were ignored. More accurate emotional processing could have been obtained if the other emotions sources were considered thus improving actual results and this would have lead to a full-fledge architecture to be deployed in a real-life pervasive environment.

As discussed later in the next section, a multimodal approach [95] is preferred so that the identification of emotions can be made accurately and each emotion detection mechanism can complement each other. Most of the existing systems developed have been considering only one or two ways of capturing emotions. However, we believe that for a pervasive system middleware to be confident enough about the emotions of its users in its environment, a multimodal approach is required, that combines facial expressions, body movements and hand gestures software components running behind cameras or using suitable sensors and software components running behind microphones and/or similar sensors to detect emotions in speech and sound.

## 4. Emotions Management

Discussion about taking emotional information as context data to better support users of a pervasive environment is one issue. What to take as emotional information is another issue. Facial expressions, speech & sound, hand gestures and body movements, as linguistic communication, are essential sources of emotions as previously discussed. However, issues such as how to detect emotions, how to represent and model the emotions, and how to incorporate emotions in system's response need to be firmly grounded before further advancing research in this area.

### 4.1 Detecting and Capturing Emotions

Facial-based emotions can be extracted or captured by image analysis and understanding [24] [25]. Ekman [88] work on coding facial expressions was based on the basic movements of facial features called Action Units (AUs). In this scheme, expressions are classified into a predetermined set of categories. Two approaches are identified, namely, the "featured-based" approach and the "region-based" approach. In the "feature-based" approach, one tries to detect and track specific features such as the corners of the mouth, eyebrows, and so on. In the "region-based" approach, facial motions are measured in certain regions on the face such as the eye/eyebrow and the mouth. In addition, we can distinguish two types of classification schemes: dynamic and static. Static classifiers (e.g., Bayesian Networks) classify each frame in a video to one of the facial expression categories based on the results of a particular video frame. Dynamic classifiers (e.g., HMM) use several video frames and perform classification by analyzing the temporal patterns of either the regions analyzed or features extracted as discussed above. They are very sensitive to appearance changes in the facial expressions of different individuals thus are more suited for person-dependent experiments [89]. Static classifiers are easier to train and in general need less training data but when used on a continuous video sequence they can be unreliable especially for frames that are not at the peak of an expression, that is, for frames that are only half-way to the complete facial expression.

Another way of capturing emotion of a user A is by monitoring the behavior, action of another user B interacting with A since B may possibly largely influence A's emotional state. For example, a student generates happiness when the lecturer praises his/her work.

Tracking of body movements (head, arms, torso, legs and others) to determine emotions is still a challenging area. Three important issues [76] emerge in articulated motion analysis: representation (joint angles or motion of all the sub-parts), computational paradigms (deterministic or probabilistic), and computation reduction. A proposal [76] based on a dynamic Markov network uses Mean Field Monte Carlo algorithms so that a set of low dimensional particle filters interact with each other to solve a high dimensional problem collaboratively. Furthermore, in body posture studies, researchers [77] used a stereo and thermal infrared video system to estimate driver posture for deployment of smart air bags. A method for recovering articulated body pose without initialization and tracking (using learning) was proposed in [78]. The authors of [79] use pose and velocity vectors to recognize body parts and detect different activities, while the use of temporal templates [80] were also investigated.

Conversation is a major channel for communicating emotion. Extracting the emotion information in conversation enables computer systems to detect emotions and capture emotional intention more accurately so as to mediate human emotions by providing instant and proper services. For exploring structure of emotion in English conversation, authors in [38] start by studying four linguistic features in English conversation such as lexical choice, syntactic form, prosody and sequential positioning.

Hand gestures play an important role in expressing emotions and allow communication with others. Most of the research and applications ranging from virtual environments [81], smart surveillance [82], and remote collaboration [83] in gesture recognition are oriented towards hand gestures. How to proceed with hand gestures recognition? First a mathematical model that considers both temporal and spatial characteristics of the hand gestures and hand need to be chosen. The model to be chosen can largely influence the nature and performance of gestures analysis. Once the model is selected, all the related model parameters need to be identified from the features that are extracted from single or multiple input streams. These parameters represent some description of the hand pose or trajectory and depend on the modeling approach used. Specific problematic areas that need to be catered for are the hand localization [84], hand tracking [85], and the selection of suitable features [86]. After the parameters are computed, the gestures represented by them need to be classified and interpreted based on the accepted model and based on some grammar rules that reflect the intended syntax of gestural commands. The grammar may also encode the interaction of gestures with other communication modes such as speech, gaze, or facial expressions. As an alternative, some authors have explored using combinations of simple 2D motion based detectors for gesture recognition [87].

## 4.2 Representing and Modeling Emotions

In this stage, data gathered from multiple sensors need to be processed, then classified based on dynamic or discriminative models and finally combined with other sensor values and results can be matched against data sets from a database to identify matching emotional patterns. Users in a pervasive environment normally express multimodal (e.g., audio and visual) communicative signals in a complementary or redundant manner [95]. Analyses of multiple input signals acquired by different sensors cannot be considered independently and cannot be combined in a context-free manner at the end of the intended analysis but, on the contrary, the input data should be processed in a joint feature space and according to a context-dependent model. Nevertheless, besides the problems of context sensing and developing context-dependent models for combining multi-sensory information, the size of the required joint feature space is an important issue. Problems include large dimensionality, differing feature formats, and time-alignment. A solution to achieve multi-sensory data fusion is to develop context-dependent versions of a suitable method such as the Bayesian inference method [96].

Emotions uncertainty is another very important issue. A pervasive system middleware needs to be versatile enough to deal with imperfect data and generate its own conclusion. To solve this problem, the time-instance versus time-scale dimension of human nonverbal communicative signals [97] can be considered. Taking into account previously observed data (time scale) with respect to the current data carried by functioning observation channels (time instance), a statistical prediction and its corresponding probability might be derived about both the information that has been lost due to inaccuracy of a particular sensor and the currently displayed action/reaction. Probabilistic graphical models, such as Bayesian networks, Hidden Markov Models and Dynamic Bayesian networks can be considered for fusing such different sources of information. These models can handle noisy features, temporal information, and missing values of features all by probabilistic inference. Despite important advances, further research is still required to investigate fusion models able to efficiently use the complementary signals captured by multiple sensors in a pervasive environment.

The HUMAINE project [7] has proposed an Emotion Annotation and Representation Language (EARL) which is an XML-based language for representing and annotating emotions. In contrast to existing markup languages, where emotion is often represented in an ad-hoc way as part of a specific language, it proposes a language aiming to be usable in a wide range of use cases, including corpus annotation as well as systems capable of recognizing or generating emotions. The proposal describes the scientific basis of emotion representations and the use case analysis through which is determined the required expressive power of the language. Core properties of the proposed language using examples from various use case scenarios are also illustrated.

EARL is thus requested to provide means for encoding the following types of information:

- *Emotion descriptor*. No single set of labels can be prescribed, because there is no agreement – neither in theory nor in application systems – on the types of emotion descriptors to use, and even less on the exact labels that should be used. EARL has to provide means for using different sets of categorical labels as well as emotion dimensions and appraisal-based descriptors of emotion.

- *Intensity* of an emotion, to be expressed in terms of numeric values or discrete labels.

- *Regulation types*, which encode a person's attempt to regulate the expression of her emotions (e.g., simulate, hide, amplify).

- *Scope* of an emotion label, which should be definable by linking it to a time span, a media object, a bit of text, a certain modality etc.

- *Combination* of multiple emotions appearing simultaneously. Both the co-occurrence of emotions as well as the type of relation between these emotions (e.g. dominant vs. secondary emotion, masking, blending) should be specified.

- *Probability* expresses the labeller's degree of confidence with the emotion label provided.

In EARL, emotion tags can be simple or complex. A simple <emotion> uses attributes to specify the category, dimensions and/or appraisals of one emotional state. Emotion tags can enclose text, link to other XML nodes, or specify a time span using start and end times to define their scope. One design principle for EARL was that simple cases should look simple. For example, annotating text with a simple "pleasure" emotion results in a simple structure:

*<emotion category="pleasure">Hello!</emotion>*

Annotating the facial expression in a picture file, for example, face12.jpg, with the category "pleasure" is simply:

*<emotion xlink:href="face12.jpg" category="pleasure"/>*

This "stand-off" annotation, using a reference attribute, can be used to refer to external files or to XML nodes in the same or a different annotation document in order to define the scope of the represented emotion. In uni-modal or multi-modal clips, such as speech or video recordings, a start and end time can be used to determine the scope:

*<emotion start="0.4" end="1.3" category="pleasure"/>*

Besides categories, it is also possible to describe a simple emotion using emotion dimensions or appraisals (Fig. 1):

```
<emotion xlink:href="face12.jpg" arousal="-0.2" valence="0.5" power="0.2"/>
<emotion xlink:href="face12.jpg" suddenness="-0.8" intrinsic_pleasantness="0.7"
goal_conduciveness="0.3" relevance_self_concerns="0.7"/>
```

Fig. 1. A simple emotion representation

EARL is designed to give users full control over the sets of categories, dimensions and/or appraisals to be used in a specific application or annotation context. Information can be added to describe various additional properties of the emotion: an emotion *intensity*; a *probability* value, which can be used to reflect the (human or machine) labeller's confidence in the emotion annotation; a number of *regulation* attributes, to indicate attempts to *suppress*, *amplify*, *attenuate* or *simulate* the state of an emotion; and a *modality*, if the annotation is to be restricted to one modality.

For example, an annotation of a face showing obviously simulated pleasure of high intensity:

*<emotion xlink:href="face22.jpg" category="pleasure" simulate="1.0" intensity="0.9"/>*

In order to clarify that it is the face modality in which a pleasure emotion is detected with moderate probability of correct prediction, we can write:

*<emotion xlink:href="face22.jpg" category="pleasure" modality="face" probability="0.5"/>*

In combination, these attributes allow for a detailed description of individual emotions that occurs in people's everyday life in different environments.

A <complex-emotion> describes one state composed of several aspects, for example because two emotions co-occur, or because of a regulation attempt, where one emotion is masked by the simulation of another one.

For example, to express that an expression could be either pleasure or friendliness, one could annotate (Fig. 2):

```
<complex-emotion xlink:href="face12.jpg">
  <emotion category="pleasure" probability="0.5"/>
  <emotion category="friendliness" probability="0.5"/>
</complex-emotion>
```

Fig. 2. Either pleasure or friendliness

The co-occurrence of a major emotion of "pleasure" with a minor emotion of "worry" can be represented as follows (Fig. 3).

```
<complex-emotion xlink:href="face12.jpg">
  <emotion category="pleasure" intensity="0.7"/>
  <emotion category="worry" intensity="0.5"/>
</complex-emotion>
```

Fig. 3. Major pleasure with minor worry

Simulated pleasure masking suppressed annoyance would be represented (Fig. 4):

```
<complex-emotion xlink:href="face12.jpg">
  <emotion category="pleasure" simulate="0.8"/>
  <emotion category="annoyance" suppress="0.5"/>
</complex-emotion>
```

Fig. 4. Pleasure masking suppressed annoyance

The numeric values for "simulate" and "suppress" indicate the amount of regulation going on, on a scale from 0 (no regulation) to 1 (strongest regulation possible). The above example corresponds to strong indications of simulation while the suppression is only halfway successful.

## 5. Conclusions and Future Works

This work discussed issues pertaining to cognition, emotion and behavioral variability in humans. A preliminary analysis of existing pervasive computing systems that take emotions into consideration based on five fundamental emotions sources, namely, facial expressions, body movements, speech and sound, hand and gestures, and linguistics, has been made. We conclude that most existing pervasive systems or smart environments middlewares have not considered a full-fledge multi-modal emotion-aware approach, though some works demonstrate that multi-modal approach is more accurate in determining users' emotional states. We also discussed the different means of detecting emotions, and emotions' representation and modeling using an XML-based language such as EARL. However, during the EARL markup language investigation, our attention was drawn to two very important issues out of the three (i.e., the two first ones) that we describe as future works below.

As future works, three important issues need to be addressed. First, a psychological and hence mathematical meaning to the regulation types that encode a person's attempt to regulate the expression of his/her emotions, for example, 'simulate', 'hide' and 'amplify', need to be developed, since values for these attributes can vary in different environments. Second, in a multi-modal emotion-aware pervasive environment, emotions detected by various sensors/devices, might result in different types of emotions, thus introducing ambiguities in determining the exact actual emotion of the user. Hence, applying the appropriate weight (referred above in EARL as *probability*) to each and every modality based on relevant contextual parameters (like, application context, environmental context and scenario context parameters) has to be firmly grounded to obtain confident values thus to determine the exact emotion of the user. Third, there is an immense field that opens in front of us: the design of smart environments that are not only *aware of human emotions*, but are also able to infer current needs from emotions and propose adapted services as a consequence.